# Deep Networks tag the location of bird vocalisations on audio spectrograms

*Lefteris Fanioudakis*
Technological Educational Institute of Crete, Department of Music Technology and Acoustics, Crete, Greece
fanioudakis.lefteris@gmail.com

*Ilyas Potamitis*
Technological Educational Institute of Crete, Department of Music Technology and Acoustics, Crete, Greece
potamitis@staff.teicrete.gr

*Abstract*—This work focuses on reliable detection and segmentation of bird vocalizations as recorded in the open field. Acoustic detection of avian sounds can be used for the automatized monitoring of multiple bird taxa and querying in long-term recordings for species of interest. These tasks are tackled in this work, by suggesting two approaches: A) First, DenseNets are applied to weekly labeled data to infer the attention map of the dataset (i.e. Salience and CAM). We push further this idea by directing attention maps to the YOLO v2 Deepnet-based, detection framework to localize bird vocalizations. B) A deep autoencoder, namely the U-net, maps the audio spectrogram of bird vocalizations to its corresponding binary mask that encircles the spectral blobs of vocalizations while suppressing other audio sources. We focus solely on procedures requiring minimum human attendance, suitable to scan massive volumes of data, in order to analyze them, evaluate insights and hypotheses and identify patterns of bird activity. Hopefully, this approach will be valuable to researchers, conservation practitioners, and decision makers that need to design policies on biodiversity issues.

I. INTRODUCTION

Birds use acoustic vocalization as a very efficient way to communicate as the sound does not require visual contact between emitting and receiving individuals, can travel over long distances, and can carry the information content under low visibility conditions, such as in dense vegetation and during night time hours [1]. In this paper we will focus only on sounds produced in the vocal organ of birds (i.e. calls and songs). The operation of autonomous remote audio recording stations and the automatic analysis of their data can assist decision making in a wide spectrum of environmental services, such as: Monitoring of range shifts of animal species due to climate change, biodiversity assessment and inventorying of an area, estimation of species richness and species abundance, assessing the status of threatened species, and alarming of specific atypical sound events related to potentially hazardous events and human activities (e.g. gun shooting) [2-3].

During the last decade the progress of bioacoustic technology is evident especially in the field of hardware development, particularly of programmable and affordable automatic recording units (ARUs). Modern models are powered by solar energy, equipped with large storage capacity, carry weather-proof normal and ultrasound microphones, and some of them are equipped with wireless transmission capabilities [4].

Pattern recognition of bird sounds has a long history and many pattern recognition approaches [5-18] have been applied to the problem of automatic bird detection and identification. This work focuses on a specific question of bird detection in audio: Is there bird activity in a recording clip? If yes, when did it happen? Can you extract segments for further examination? Although the approaches we describe are directly expandable to more refining questions, in this work we investigate bird activity in general and we are indifferent to species' identity. That is, we present a generic bird activity detector of vocalizations. The described approaches set a bounding box in the time-frequency spectrum corresponding to bird vocalizations, therefore allows time-stamping, extraction and retrieval of sound snippets. Once trained, they are very fast in execution; require only minimal human attendance during training and none once operational.

The reported literature on the application of Deep learning networks on bird audio recordings was until recently sparse [17-18]. This work introduces different types of deep learning networks to this particular task. Our novelties are as follows:

A) We elaborate on the line of though initially reported in [18] and we introduce distinct improvements, namely: By using pretrained DenseNets on Imagnet and adapting the models on spectrograms we reach higher scores than these reported in the birds detection challenge [18-19]. Subsequently, we derive the Salience map of the training and validation set. A second YOLO v2 architecture is trained on the Salience maps to predict spectral segments containing birds vocalisations. B) a U-net [20] autoencoder is used to detect bird vocalizations by mapping the spectrogram's blobs to binary masks.

I. DEEP NETS AND AND THE SPECTROGRAM

Bird calls usually refer to simple frequency patterns of short monosyllabic sounds. While all birds emit calls, although with different variability and frequency, only some birds also produce songs. In difference to calls, songs are longer, acoustically more complex, and often have a modular structure [1-3]. The spectrogram –also called Short-time Fourier transform- is the outcome of a number of processing stages imposed on audio. The sampled data in the time domain stored in the ARUs, are decomposed into overlapping data chunks that are windowed. Each chunk is subsequently Fourier transformed and the magnitude of the frequency spectrum of each data-chunk is derived. Each spectrum vector corresponds to a vertical line in the image; a measurement of magnitude versus frequency for a specific moment in time. These spectrum vectors are placed side by side to form the spectrogram image. An audio scene can be treated as an image through its spectrogram. Acoustic events appear as localised spectral blobs/patches on a two-dimensional matrix (see Fig. 1). The structure of these blobs constitutes the acoustic signature of the sound and is used as a biometric queue to reveal evidence of identity of the source is several bioacoustics applications.

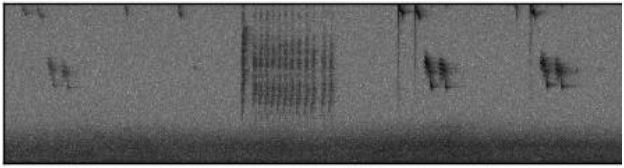

Fig. 1. Spectrogram of bird calls in the presence of strong wind. Audio events stand out as patches of intense coloring. We remove axis to emphasize the notion of a spectrogram as a canvas of spectral blobs corresponding to birds' vocalizations.

*A. DenseNet and saliency maps*

The Dense Convolutional Network (DenseNet), connects each layer to every other layer in a feed-forward fashion [18, 21]. We used 121 and 169 DenseNets with pre-trained weights trained on Imagenet database (121, 169 denotes the depth of the ImageNet models). Then we adapted the weights on spectrograms that are copied to the RGB channels of the input. The spectrograms use a 512 hamming window and FFT and when applied to the recordings of [22] return 256x624 spectrograms where 624 is the number of audio frames. Subsequently, all spectrograms are reshaped to 224x224 to become compatible with what the DenseNet expects as a figure's input size.

*B. Saliency maps and bird vocalisations*

Bird vocalization recordings with exact boundaries are costly and rare for large datasets. It is easy to annotate a recording as having a bird vocalization or not based on visual inspection of its spectrogram. It is costly however to derive bounding boxes for all vocalizations inside. On the contrary weakly labeled data are abundant (see e.g. the Xeno-canto database at http://www.xeno-canto.org/ ). Weakly labelled in the context of this work means that a recording is labeled as having a bird sound or not but there are no other metadata on where is the bird sound exactly located within the recording. Predicting the exact location of the vocalization allows different kind of measurements to be derived e.g. bird activity per unit time, extraction of the repertoire of vocalizations, recognition of different species. In this work, as in [18], we use the Salience map as a by-product of Deep-nets that allows us to localize the vocalizations. The Salience map allows as to have a glimpse on where exactly the deep net basis its decision to classify a recording as having or not a bird vocalization. Thus, implicitly, the Salience map tags the spectrogram with the correct localization of the vocalizations (see Fig 2). Once we derive the Salience map of the part of the available database having a positive label for birds, we apply bounding boxes on the saliency blobs and then we apply YOLO v2 to derive bounding boxes for the part of the test set classified by the DenseNet as having a bird.

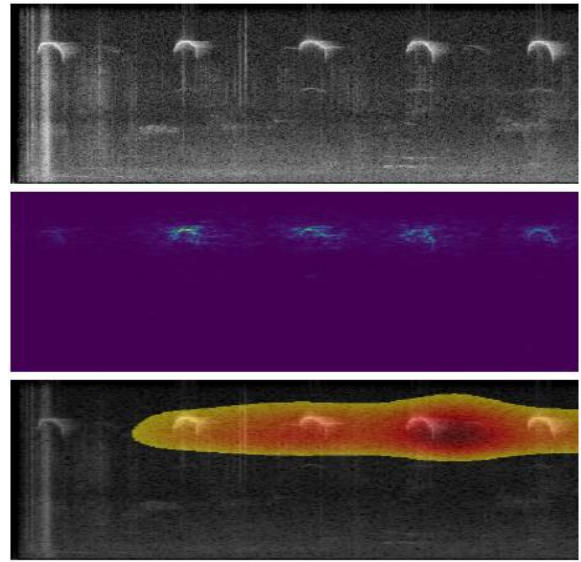

Fig. 2. Spectrogram of file '0abeb112-2bb9-4b2a-804b.wav' (TOP) and its corresponding Saliency Map (MIDDLE) and Class Activation Map (CAM) (BOTTOM).

Two different types of attention mapping generated automatically a) the guided-backprop Saliency Map, and b) the gradient Class Activation Map (grad-CAM) (see Fig. 2 and Acknowledgments). The CAM prefers to group spectral blobs belonging to birds instead of segmenting phrases like the Saliency map does, which may have affected the object detection training as the Table II shows in the Results section. Attention mapping is a valid procedure to extract birds' vocalizations by itself. To see if a DeepNet can mitigate the errors the attention mapping produce we direct the spectrogram patches that correspond to the attention maps to state of the art object detection technique to predict a second set of bounding boxes different to the attention maps. Thus, we ended up using YOLO v2 object detector which has demonstrated better performance in our task among state of the art detection techniques such as SSD, and FASTER R-CNN. YOLO v2 uses a Deep network architecture for both classification and localization of the object, using bounding box regression and classification. We edited network's configuration file to correspond to our specific class and files and we left the resolution at 416x416. By using the extracted coordinates of our bounding boxes from the attention blobs, we trained YOLO v2 object detector with the pre-trained ImageNet weights of Darknet19 448x448 which is based on the Extraction model (See Appendix). The benefit of using YOLO instead of attention maps solely are: a) better localized bounding boxes on vocalizations than attention maps, b) YOLO is very fast in predicting bounding boxes whereas attention maps take much time to be created.

*C. U-nets and spectrogram segmentation*

The U-net architecture [20] consists of a contracting path to capture context around the blobs that ends to a bottleneck and subsequently, a symmetric expanding path that enables the determination of a binary mask imposed on the picture that finally allows, in our case, the localization of spectral blobs belonging to birds vocalizations. In this work, we use a modified version of [6] to extract automatically the mask of the spectrogram of a bird recording (see Fig. 3-bottom). The training set is composed of spectrogram figures of bird recordings as well as recordings void of any bird activity and their corresponding binary masks. Recordings having audio events other than bird vocalizations are mapped to zero maps. During training, the spectrogram which is a 2D representation is presented as input, and the mask (e.g. Fig. 3 BOTTOM) is presented as output, whereas the network learns the mapping in-between them.

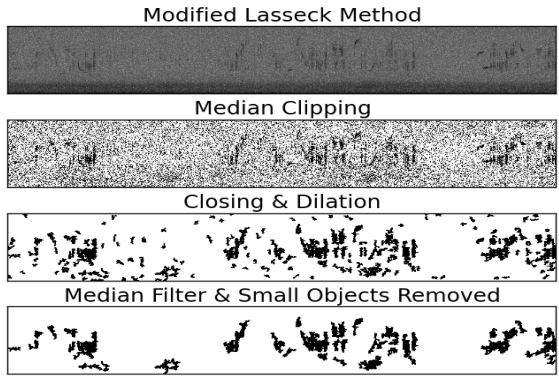

Fig. 3. Segmentation based on a modified method of Lasseck in [6], The spectrogram of a bird recording is led through the processing stages of median clipping, morphological operations involving closing & dilation and finally median filtering to a binary mask of presence/absence of audio activity in a spectral patch.

## II. RESULTS

The Dataset described in [19] consists of over 17,000 ten-second audio recordings and their associated binary, hand-labelled tags corresponding to the presence/absence of a bird sound in each clip. The recordings include vocalizations of various bird species recorded in the field and recordings containing acoustic events other than bird sounds.

*A. Densnet Bird detection Results*

In [18-19] one needs to classify a recording of either having or not a bird vocalization (i.e. a binary decision). In Table I we gather comparative results of DenseNet versions on the same random holdout set (20%). All models are pre-trained on ImageNet and adapted for 50 epochs on the training corpus of [19]. Note than the main difference of our results and [18] is the use of pre-trained weights. In Table I, mean subtraction stand for subtracting the mean value from each frequency channel of the spectrogram before feeding it to the deep net. 'Reconstructed Spectrogram' stands for making a spectrogram out of a Mel-filterbank spectrogram; a process that smooths out the spectrogram. We used different versions of smoothed and enhanced spectrograms instead of copying 3 identical versions of the spectrogram to the input tensor but, unfortunately, we did not observe any distinct gain.

| Model and Input | ACC(%) | AUC (%) |
|---|---|---|
| 121-DenseNet, raw spectrogram | 87.8 | 93.53 |
| 121-DenseNet, spectrogram, mean subtraction | **88.94%** | **94.76%** |
| 121-DenseNet, raw spectrogram reconstructed Spectrogram | 87.89% | 93.53% |
| 169-DenseNet, spectrogram, mean subtraction | 88.75% | 94.61% |

Table I. Comparative results on the same random holdout set (20%).

Accuracy classification score computes a subset accuracy: the set of labels predicted that match the corresponding set of true labels. AUC is the Area Under the Receiver Operating Characteristic Curve (ROC AUC) and needs both the validation labels as well as the prediction probabilities. This measure measures how confident is the classifier about its decisions.

*B. Densnet Bird-vocalizations, segmentation Results*

Our labeled data consisted of 7980 samples, of which 20% randomly selected, have been holdout. Note that we use the part of the training set tagged as having a bird vocalization to extract attention maps. Training took place for about 6000 iterations, respectively for both attention map cases, which was sufficient for comparative results (see Table II).
As an evaluation metric we use Intersection over Union for object detection (IOU) using the ground-truth bounding boxes (extracted b-boxes from attention maps in our case) and the predicted bounding boxes from our trained YOLO v2 model, at the holdout set (see Fig. 4).

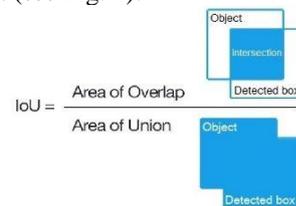

Fig. 4. A graphical explanation of the IOU metric. Dividing the area of overlap between the bounding boxes by the area of union gives us the accuracy.

| Attention Map | Best Iter. | IOU (%) |
|---|---|---|
| Gradient Class Activation | 5400 | 65.62% |
| Guided Backprop Saliency | 4600 | 66.64% |

Table II. Comparative YOLO v2 intersection over union results on the same random holdout set (20%).

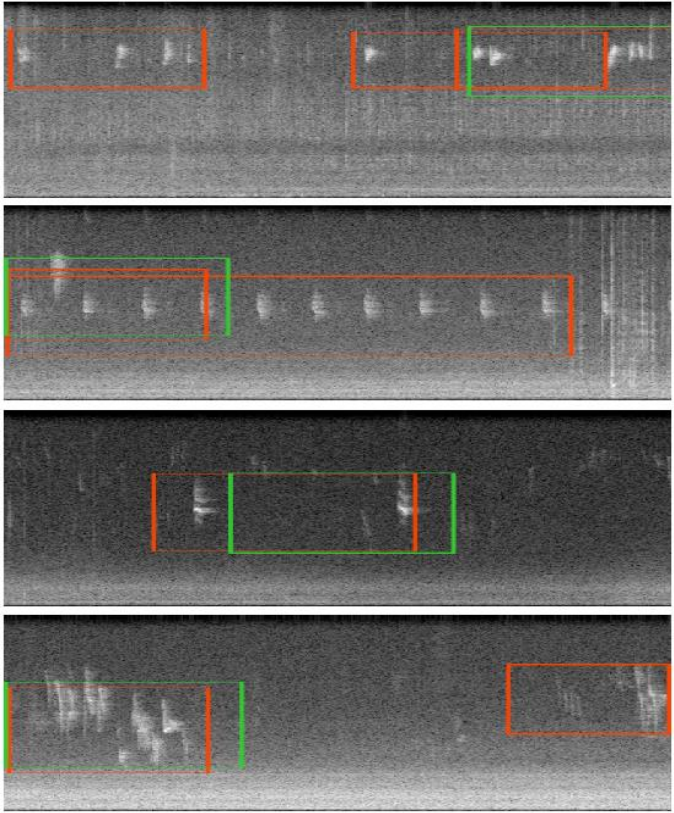

Fig. 5. YOLO v2 predictions from gradient Class Activation trained model on holdout spectrograms. Green bounding boxes represents our ground-truth attention maps, and red the predicted one.

One can see some typical detection and segmentation results in Fig 5. The trained blob detector can localize better and faster than the attention map itself. As we mentioned before, prediction speeds differ greatly from those of the attention maps alone. The YOLO v2 approach is quite accurate in detecting bird vocalization in complex acoustic environments while disregarding spectral blobs originating from acoustic interferences.

### C. U-net Bird-vocalization segmentation Results

Note that, the ground truth of vocalization masks is missing and these are approximated by the masks derived automatically by the method in [6]. The Lasseck method derives blindly the masks of spectral blobs, and, therefore, can be only partially correct. The accuracy of the bird vocalizations masks varies depending on the noise present in a recording. The training based on partly accurate masks is improved by including in the training process recordings having environmental sounds but, otherwise, being empty of bird vocalizations. As the latter recordings are mapped to zero masks, the network improves over time to correct, -to a certain extent that is- the effect of using partially correct training masks. We trained the U-net detection framework in terms of the mean Dice coefficient loss function. The Dice coefficient can be used to compare the pixel-wise agreement between a predicted segmentation and its corresponding ground truth. The formula is given by:

$$\frac{2*|X \cap Y|}{|X|+|Y|}$$

Where, X is the predicted set of pixels and Y is the ground truth. The Dice coefficient is the quotient of similarity and ranges between 0 and 1. It can be viewed as a similarity measure over sets. The loss function is just the minus of the Dice coefficient with the additions of a smoothing factor inserted in the denominator. The score in Table III is the mean of the Dice coefficients of images in the evaluation set.

| U-net | Dice Coef. | Train time (h) | Pred. time/im (s) |
|---|---|---|---|
| Simple U-net | 0.71 | 5 | 0.06 |
| Enlarged U-net | 0.74 | 7 | 0.16 |
| Inception blocks | 0.65 | 10 | 0.79 |

Table III. All U-net versions are trained with 60 epochs on the same dataset. The Inception block converges slower due to the small batch-size necessary to avoid memory overflow but finally achieves better results after a large number of epochs.

In Fig. 6 one can see an illustrative example of the U-net predicting a mask over bird vocalizations for a noisy recording.

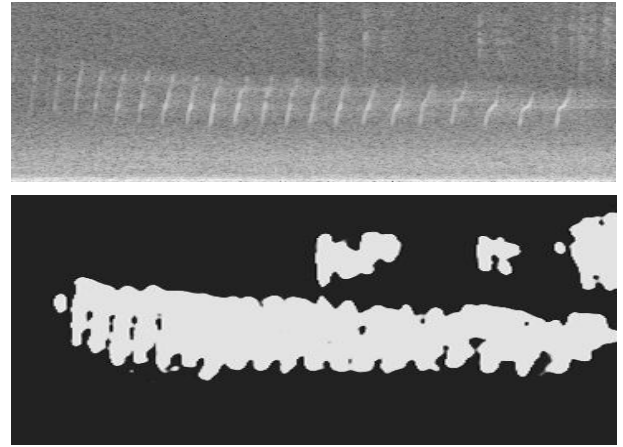

Fig. 6. An audio-scene with bird activity in the presence of strong wind noise. (TOP) Spectrogram, (BOTTOM) Predicted binary mask of bird vocalizations.

### III. DISCUSSION

The process of manual tagging the exact locations of bird vocalizations in a recording is laborious and problematic when it needs to be performed for thousands of recordings. Our aim is to automate the procedure of tagging the locations of bird vocalizations in the spectrogram. We have identified two ways: a) We use the Salience/CAM map of a DenseNet to automatically tag the spectrogram patches on which they based their decision to classify a whole spectrogram as having a bird vocalization or not. Attention maps implicitly tag the location of the vocalizations and therefore the dataset is automatically annotated. We tried a refinement of this approach by directing spectrogram patches as tagged by attention maps to be handled by the YOLO v2 framework to derive refined bounding boxes of spectral patches belonging to bird vocalizations.

b) The Lasseck method [6] is used to derive spectral blobs in the spectrum of recordings. This method is blind to whether the spectral blobs originated from a singing bird or from another audio source e.g. interference. Again, the U-net that predicts vocalization masks improves itself and finally gets fine-tuned by mapping recordings with no vocalizations to zero masks.

## Acknowledgment

We gratefully acknowledge the support of NVIDIA Corporation with the donation of a TITAN-X GPU partly used for this research. For all DenseNets implementations we used the Keras 2 framework on top of the TensorFlow backend. This work was partially supported by the European Commission FP7, under grant agreement n°605073, project ENTOMATIC. We made use of parts of the following software and instructions (6/11/2017):

https://github.com/fchollet/keras
https://arxiv.org/abs/1506.02640
https://pjreddie.com/darknet/yolo/
https://arxiv.org/pdf/1412.6806.pdf
https://github.com/experiencor/deep-viz-keras
https://arxiv.org/pdf/1610.02391.pdf
https://github.com/raghakot/keras-vis

Fig. 5 made use of the recordings 1ab6b64f-8210-4752-a9f9.wav, 8b1c7003-723d-40b0-b5d2.wav, e222d3aa-588c-4e85-8e95.wav, f315b8d0-31fc-4f7a-8352.wav [19].